\documentclass[twocolumn]{aastex63}
\usepackage{graphics}
\usepackage{color}
\usepackage{topcapt}
\usepackage{amsmath,amstext}
\usepackage{empheq} 
\newcommand{\msun}{M$_{\odot}$}
\newcommand{\mstar}{$M_*$}

\newcommand{\teighty}{$\tau_{80}$}
\newcommand{\tninety}{$\tau_{90}$}

\newcommand{\hi}{H{\sc i}}

\newcommand{\peg}{Pegasus~W}
\newcommand{\leom}{Leo~M}
\newcommand{\leok}{Leo~K}
\definecolor{amaranth}{rgb}{0.9, 0.17, 0.31}
\shortauthors{McQuinn et al.}
\shorttitle{\leom\ and \leok}

\begin{document}
\title{Discovery and Characterization of Two Ultra Faint Dwarfs Outside the Halo of the Milky Way: Leo M and Leo K}

\author[0000-0001-5538-2614]{Kristen.~B.~W. McQuinn}
\affiliation{Department of Physics and Astronomy, Rutgers, The State University of New Jersey, 136 Frelinghuysen Rd, Piscataway, NJ 08854, USA}
\email{kristen.mcquinn@rutgers.edu}

\author[0000-0002-1200-0820]{Yao-Yuan Mao}
\affiliation{Department of Physics and Astronomy, University of Utah, 115 South 1400 East, Salt Lake City, UT 84112, USA}

\author[0000-0002-9599-310X]{Erik J. Tollerud}
\affiliation{Space Telescope Science Institute, 3700 San Martin Drive, Baltimore, MD 21218, USA}

\author[0000-0002-2970-7435]{Roger~E. Cohen}
\affiliation{Department of Physics and Astronomy, Rutgers, The State University of New Jersey, 136 Frelinghuysen Rd, Piscataway, NJ 08854, USA}

\author[0000-0003-3408-3871]{David Shih}
\affiliation{Department of Physics and Astronomy, Rutgers, The State University of New Jersey, 136 Frelinghuysen Rd, Piscataway, NJ 08854, USA}

\author[0000-0003-1109-3460]{Matthew~R. Buckley}
\affiliation{Department of Physics and Astronomy, Rutgers, The State University of New Jersey, 136 Frelinghuysen Rd, Piscataway, NJ 08854, USA}

\author[0000-0001-8416-4093]{Andrew E. Dolphin}
\affiliation{Raytheon Technologies, 1151 E. Hermans Road, Tucson, AZ 85756, USA}
\affiliation{University of Arizona, Steward Observatory, 933 North Cherry Avenue, Tucson, AZ 85721, USA}

\begin{abstract}
We report the discovery of two ultra-faint dwarf galaxies, \leom\ and \leok, that lie outside the halo of the Milky Way. Using Hubble Space Telescope imaging of the resolved stars, we create  color-magnitude diagrams reaching the old main sequence turn-off of each system and (i) fit for structural parameters of the galaxies; (ii) measure their distances using the luminosity of the Horizontal Branch stars; (iii) estimate integrated magnitudes and stellar masses; and (iv) reconstruct the star formation histories. Based on their location in the Local Group, neither galaxy is currently within the halo of the Milky Way, although \leok\ is located $\sim26$ kpc from the low-mass galaxy Leo~T and these two systems may have had a past interaction. \leom\ and \leok\ have stellar masses of $1.8^{+0.3}_{-0.2}\times10^4$ \msun\ and $1.2\pm0.2\times10^4$ \msun, and were quenched $10.6^{+2.2}_{-1.1}$ Gyr and $12.8^{+0.1}_{-4.2}$ Gyr ago, respectively. Given that the galaxies are at farther distances from the MW, it is unlikely that they were quenched by environmental processing. Instead, given their low stellar masses, their early quenching timescales are consistent with the scenario that a combination of reionization and stellar feedback shut-down star formation at early cosmic times. 
\end{abstract} 

 
\section{Introduction}\label{sec:intro}
The smallest galaxies in the universe contain negligible baryonic mass and are aptly named `ultra-faint dwarf' galaxies ($M_*<10^5$ \msun; $M_V> -7.7$ mag) \citep[e.g.,][]{Simon2019}. Archeological studies show these systems are the oldest, most metal-poor, and most dark matter-dominated galaxies known \citep[e.g.,][]{Munoz2006, Martin2007, Simon2007, Kirby2008, Kirby2011, Geha2013, Brown2014}, making them unique probes of galaxy formation and evolution theories.

Our prevailing theoretical framework suggests that star formation and the stellar mass assembly of ultra-faint dwarfs are halted at early cosmic times by reionization with an assist by stellar feedback. In this scenario, UV photons in the epoch of reionization inhibit the accretion and collapse of gas into stars in  small  halos, while stellar feedback from inside the systems simultaneously heats and ejects gas. Simulations concur that the critical mass for such rapid quenching of ultra-faint dwarfs is \mstar $\sim10^5$\msun\ in Local Group-like environments and that the reionization-driven quenching occurs early \citep[e.g.,][]{Wetzel2015, Fillingham2015, Benitez-Llambay2015, Fillingham2016, RodriguezWimberly2019, Katz2020}. This theoretical framework is supported by detailed observational constraints of ultra-faint dwarf galaxies that are satellites of the more massive hosts in the Local Group, namely the satellites of the Milky Way (MW), M31, and the Magellanic Clouds. The observational results show that the ultra-faint dwarfs have indeed been quenched at early times \citep[with some low-level star formation occurring at intermediate times; e.g.,][]{McConnachie2009, Brown2014, Bechtol2015, Drlica-Wagner2015, Koposov2015, Collins2022, Sacchi2021, Savino2023} and this early quenching is generally attributed to reionization. 

However, a low-mass halo can also be quenched by environmental forces (e.g., ram pressure and tidal stripping) and distinguishing between the effects of environment and reionization quenching at this low mass scale ($\lesssim 10^5$ \msun) is not entirely straightfoward. There are observational hints that the quenching fraction is higher for satellites with present-day distances closer to the host \citep[among $\sim$400 satellite galaxies around 30 Local Volume hosts in the ELVES survey;][]{Greene2023}. Simulations suggest that star formation in ultra-faint dwarfs can be quenched by environmental processing at distances as far as $\sim2$ $R_\text{virial}$ \citep[e.g.,][]{Fillingham2018}. Simulated FIRE-2 satellites down to Log M$_{\star}$/M$_{\odot}$=5 show stronger ram pressure stripping in paired (i.e., Local Group-like) versus single host halos due to increased gas density \citep{Samuel2022a}. In addition, ultra-faint dwarfs in both dark-matter-only \citep[e.g.][]{Fillingham2019} and full zoom-in hydrodynamical cosmological simulations quench well before infall to a massive galaxy, in some cases before reaching even 3$R_{vir}$ \citep{Applebaum2021}.

Constraining the role of environment on the early evolution of ultra-faint dwarfs is challenging as it depends critically on understanding a galaxy's location relative to other systems $\sim10$ Gyr ago. Placing such useful constraints on the orbital histories of ultra-faint dwarfs is difficult, particularly given that detailed modeling shows that non-static potentials are crucial to understanding the orbital histories of infalling ultra-faint dwarfs \citep{Miyoshi2021, Armstrong2021}. 

What is needed to reduce the ambiguity of environment effects is a comparable sample of galaxies that are at farther distances from massive host at the present-day, but, to date, we know very little about the properties and early mass assembly of very low-mass galaxies that are not currently found within the halo of their host. Only recently were SFHs of ultra faint dwarfs located outside the virial radius of a massive galaxy reported, namely for Eridanus~II and Pegasus~W \citep[e.g.,][]{Simon2021, McQuinn2023}. Interestingly, while the stars in Eridanus~II are uniformly old ($\sim13$ Gyr), \peg\ shows an extended SFH over $\sim7$ Gyr and was clearly not quenched by reionization despite having only \mstar\ $=6.5^{+1.1}_{-1.5} \times 10^4$ \msun. While just one system, the SFH of \peg\ suggests that either the mass scale for quenching galaxies is lower than previously thought, quenching by reionization depends more strongly on the distance to a massive galaxy at early times, or there are other factors to consider.

To more fully explore the formation and evolution of ultra-faint dwarfs, including the impact of environment on their properties, we are working to find and characterize additional ultra-faint dwarfs located in different environments. Here, we report the discovery of two more ultra-faint dwarfs, \leom\ and \leok, that reside outside the virial radius of a massive galaxy, and we characterize their properties using follow-up Hubble Space Telescope imaging of their resolved stars. 

\begin{table}
\begin{center}
\caption{Properties of \leom\ and \leok}
\label{tab:properties}
\end{center}
\begin{center}
\vspace{-15pt}
\begin{tabular}{lrr}
\hline 
\hline 
					& \leom\ 			& \leok\ \\
Property				& Value 			& Value \\
\hline
RA (J2000) 				& 166.3393$^{+0.0007}_{-0.0004}$$^{\circ}$ & 141.0255$^{+0.0004}_{-0.0003}$$^{\circ}$\\
Dec (J2000)				& 25.34529$^{+0.0007}_{-0.0004}$$^{\circ}$ & 16.5105$^{+0.0004}_{-0.0002}$$^{\circ}$ \\
P.A. $\theta$ ($^{\circ}$ E of N) & $-51^{+9}_{-7}$ & $-69^{+16}_{-12}$ \\
ellipticity ($\epsilon= 1 - \frac{b}{a}$) & 0.61$\pm0.01$ & 0.41$\pm0.01$ \\
S{\'e}rsic n parameter  & $0.88^{+0.09}_{-0.08}$ & $1.17^{+0.03}_{_0.04}$ \\
$f_b$           		& 0.002$^{+0.004}_{-0.002+}$  &  0.099$^{+0.049}_{-0.045}$ \\
$r_h$ (\arcsec)			& 59.8$^{+2.3}_{-2.1}$ 	          & 38.2$^{+3.8}_{-3.5}$\\
$r_h$ (pc)				& 133$^{+8}_{-7}$	              & 80$^{+15}_{-41}$ \\
$M_V$ (mag)				& $-5.77^{+0.15}_{-0.16}$ 	      & $-4.86^{+0.83}_{-0.29}$ \\
\mstar\ (\msun)			& $1.8^{+0.3}_{-0.2}\times 10^4$  & $1.2\pm0.2\times10^4$\\
HB m$_{V, 0}$ (mag)		& $23.79^{+0.41}_{-0.04}$ 	      & 23.65$^{+0.12}_{-0.55}$ \\
$\mu$	(mag)			& 23.31$^{+0.10}_{-0.09}$ 	      & 23.19$^{+0.08}_{-0.64}$\\
Distance (kpc)			& 459$^{+21}_{-18}$		          & 434$^{+17}_{-127}$ \\
$[$M/H$]$ (dex)	 		& $-1.9\pm0.1$ 		              & $-1.9\pm0.1$\\
\teighty\ (Gyr)   	    & $10.6^{+2.2}_{-1.1}$ 	& $12.8^{+0.1}_{-4.2}$ \\ 
$A_V$ (mag)				& 0.045  			& 0.073 \\
$A_{F606W}$ (mag)		& 0.040 			& 0.066 \\
$A_{F814W}$ (mag)		& 0.025  			& 0.041 \\
F060W 50\% (mag)		& 27.5				& 27.4 \\
F814W 50\% (mag)		& 26.5				& 26.4 \\
\hline
\hline              
\end{tabular}
\end{center}
\tablecomments{The properties of \leom\ and \leok\ were measured in this work, with the exception of the foreground extinction which is  from \citet{Schlafly2011}. $\mu$ represents the distance modulus. The 50\% values are the 50\% completeness limits per filter as measured from ASTs.}
\end{table}

\begin{figure*}
\begin{center}
\includegraphics[width=0.43\textwidth]{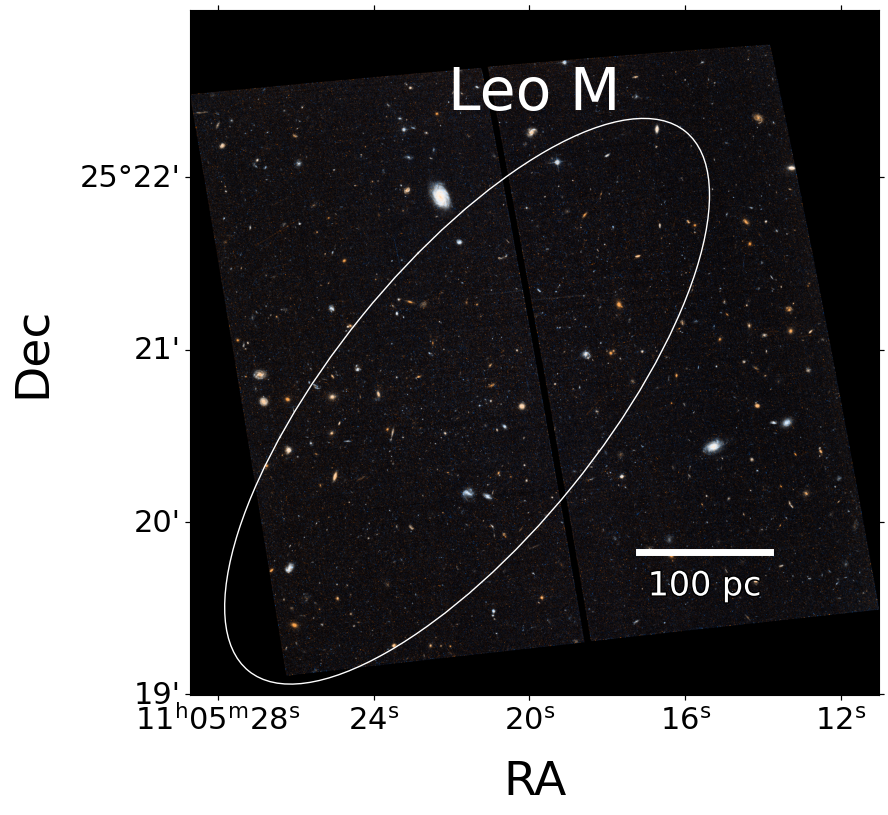}
\includegraphics[width=0.44\textwidth]{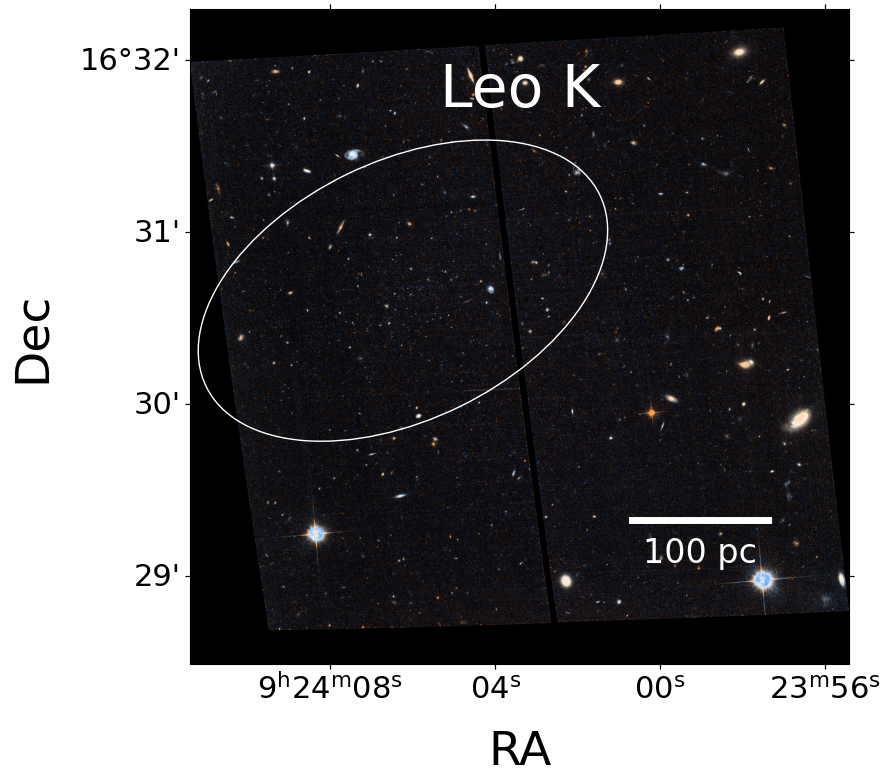}
\end{center}
\caption{Color images for \leom\ (left) and \leok\ (right) based on the ACS data with ellipses encircling the stellar component of the galaxies out to $2~r_h$ based on the best-fitting structural parameters. The galaxies are barely visible in the images, but careful inspection reveals a population of stars distinct from the surrounding field regions. All images were created using F606W for red, and average of F606W and F814W images for green, and F814W for blue.}
\label{fig:image}
\end{figure*}

The paper is organized as follows. In Section~\ref{sec:data}, we describe the observations, the data processing steps, and measurements of the structural parameters of the galaxies. In Section~\ref{sec:cmd}, we present the color-magnitude diagram of the final stellar catalogs, measure the distances to the galaxies based on the horizontal branch stars, and determine the location of the galaxies within the Local Group relative to the MW and its satellites. In Section~\ref{sec:sfh}, we derive the star formation histories, measure the stellar masses and quenching timescales. In Section~\ref{sec:discuss} we discuss our findings in the context of a broader population of ultra-faint dwarfs and in Section~\ref{sec:conclusions} we summarize our conclusions. 

\section{Observations and Data Processing}\label{sec:data}
The galaxies were identified in the DESI Legacy Imaging Surveys data as overdensities in the photometric stellar catalog, similar to the ultra-faint dwarf galaxy \peg\ \citep{McQuinn2023}. The galaxy names were based on adopting the constellation in which they reside plus a letter, similar to other low-mass galaxies at farther distances from the MW (e.g., Leo~A, Leo~T, and Leo~P). {\em HST} imaging of the galaxies was obtained as part of the HST-GO-16916 program (PI McQuinn),  which also acquired imaging of the ultra-faint dwarf galaxy \peg\ \citep{McQuinn2023}. The observation set-up for all three systems followed the same strategy and we applied a uniform approach to processing the data. Here, we briefly summarize the data and analysis and refer the interested reader to the more detailed description provided in \citet{McQuinn2023} for \peg. 

\subsection{Images and Photometry}
Both galaxies were observed for 1 orbit with Advanced Camera for Surveys (ACS) instrument \citep{Ford1998} as the primary instrument and the Wide Field Camera 3 (WFC3) UVIS instrument in parallel. The purpose of the parallel imaging is to assess the level of foreground and background contaminant sources in a nearby area of the sky. The single HST orbit per target was split between the F606W and F814W filters on each instrument and included 2 exposures per filter with a small dither between exposures ({\tt acs wfc dither line pattern \#14}). Total exposure times were 1020s in ACS F606W, 1045s in ACS F814W, and 1140s in both WFC3 F606W and F814W. All the {\it HST} data used in this paper can be found in MAST: \dataset[10.17909/x8qj-bn51]{http://dx.doi.org/10.17909/x8qj-bn51}.

Figure~\ref{fig:image} presents combined color images of \leom\ and \leok\ from the ACS data created using the {\tt flc.fits} and the {\em HST} {\tt drizzlepac v3.0} python package \citep{Hack2013, Avila2015}. The galaxies are barely visible by eye but careful inspection of the images reveals a population of point sources that make up each galaxy.

\begin{figure*}
\begin{center}
\includegraphics[width=0.47\textwidth]{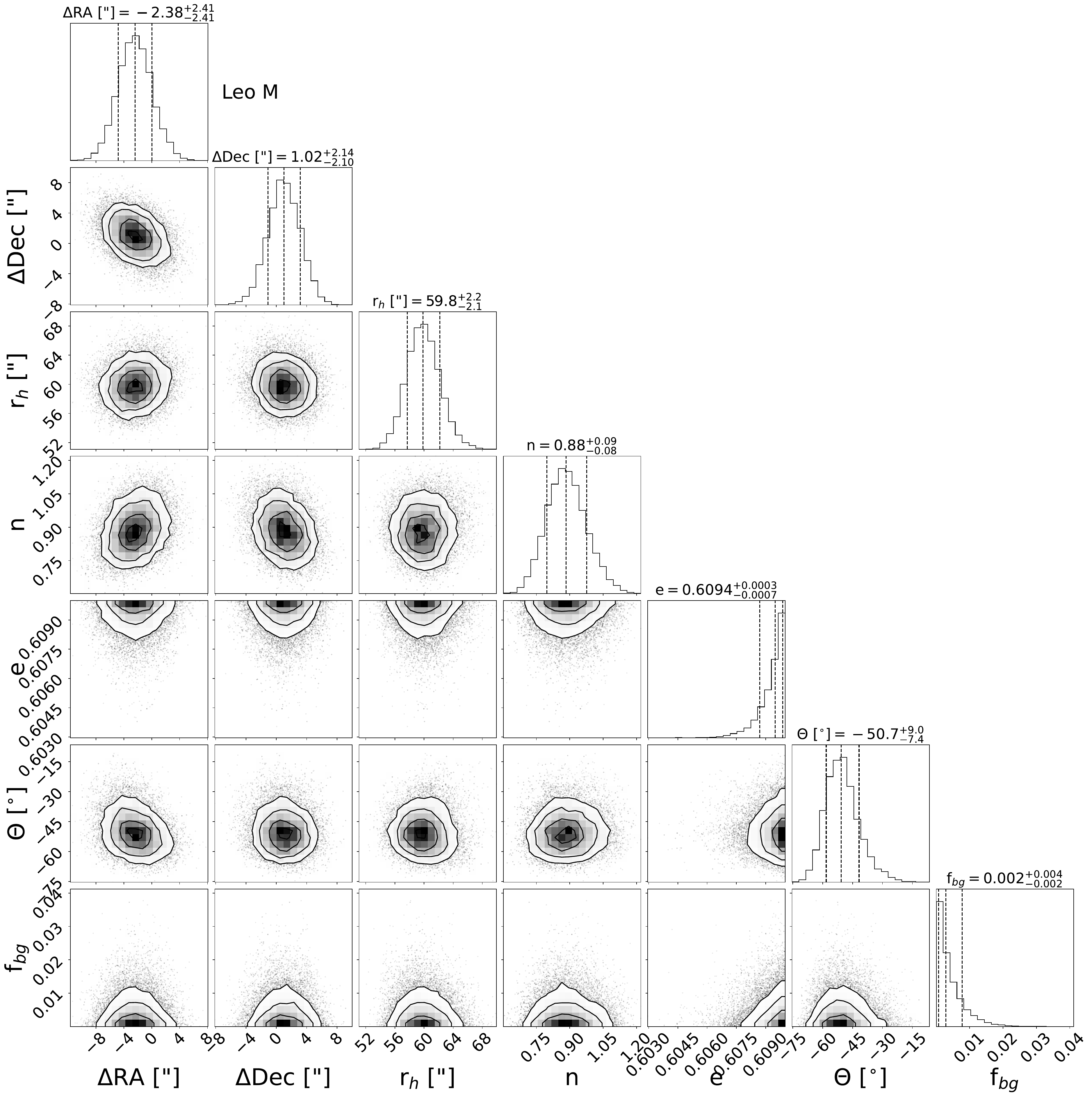} 
\includegraphics[width=0.47\textwidth]{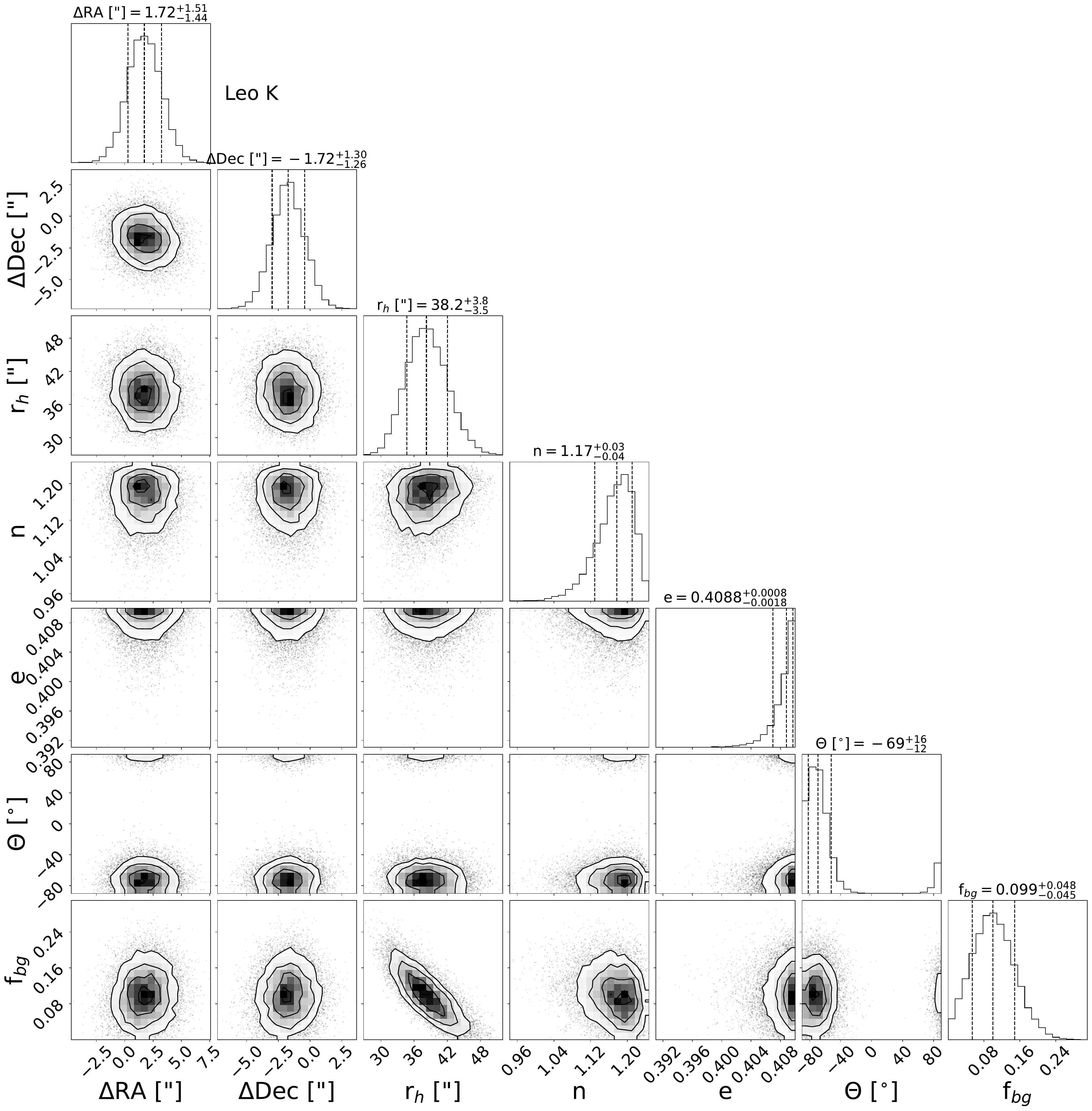} 
\end{center}
\caption{Full posterior distributions of the structural parameters for \leom\ (left) and \leok\ (right) using an equally-weighted resampling of the nested sampling posterior sample points (see text). The black contour lines correspond to 1,2 and 3$\sigma$. The $\Delta$RA and $\Delta$Dec values are relative to an initial guess for the center of the galaxies based on the distribution of well-recovered sources in the photometric catalog. $f_{bg}$ represents the potential background density in fractional form. Best-fitting values are provided in the plot and are also listed in Table~\ref{tab:properties}.}
\label{fig:corner}
\end{figure*}

\begin{figure*}
\begin{center}
\includegraphics[width=0.47\textwidth]{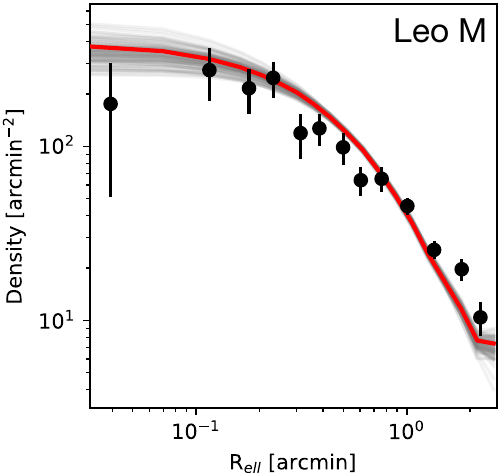} \includegraphics[width=0.47\textwidth]{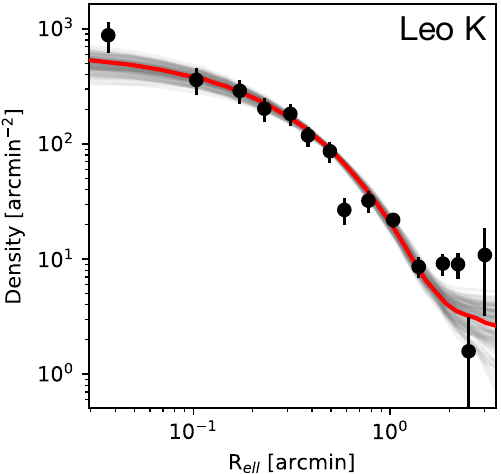} 
\end{center}
\caption{Binned radial density profile of the observed stars in \leom\ (left panel; black points) and \leok\ (right two panel; black points); errorbars are based on Poissonian uncertainties. The exponential profiles from our maximum likelihood analyses are overlaid in red and the grey lines represent 100 random draws from the posterior distributions of the profile parameters.}
\label{fig:densprof}
\end{figure*}

Point spread function (PSF) photometry was performed on the {\tt flc.fits} files using the software package {\tt DOLPHOT} \citep{Dolphin2000, Dolphin2016} and the HST module-specific PSF libraries. We used the same photometry parameters determined through extensive tests and described in \citet{Williams2014, Williams2021}. We filtered the full {\tt DOLPHOT} output for well-recovered stars that met the following criteria: output error flag $<4$, object type $\leq2$, signal-to-noise ratio $\geq5$ in both filters, sharp$_{F606W}^2 +$ sharp$_{F814W}^2 < 0.075$, and crowd$_{F606W} + $ crowd$_{F814W} < 0.1$. {\tt DOLPHOT} was also used to perform $\sim300$k artificial star tests on the images in the region of the galaxies, providing a measure of completeness and observational uncertainties. We applied the same quality cuts to the photometrically recovered artificial stars, resulting in 50\% completeness limits of 27.5 and 27.4 mag in the F606W filter and 26.5 and 26.4 mag in the F814W filter for \leom\ and \leok, respectively.

\subsection{Structural Parameters}\label{sec:structure}
In order to better characterize the galaxies' properties, we determined the geometry and structural parameters for both systems. We measured structural parameters using a Bayesian inference approach inspired by similar analyses like \citet{Martin2008, Martin2016}. We started from the spatial distribution of well-recovered sources in the ACS data that were above the 50\% completeness limits in each filter (F606W$<27.5$, F814W$<26.5$ for \leom; F606W$<27.4$, F814W$<26.4$ for \leok). To these, we fit a model with two components: a S{\'e}rsic profile \citep{sersic}, and a uniform density. Because S{\'e}rsic profiles have long tails for large $n$, this profile was normalized numerically over the convex hull containing all of the stars and the uniform background was normalized over the same area.

We then used the {\tt dynesty} nested sampling code \citep{dynesty1} to infer the parameters for this model. This requires assumed priors for all the parameters in this model.  We assumed uniform priors on the center of the S{\'e}rsic profile, covering the middle 80\% of the source catalog to avoid edge effects.  We used a trucated "scale-free" prior (sometimes called the ``Jeffrey's Prior'') for the half-light radius of the S{\'e}rsic profile, truncated to cover 7.2 to 54 arcsec, as this covers the peak of the posterior while not wasting large amounts of execution time on unphysically small or large profiles.  

For the position angle and fraction of sources in the background distribution, we adopted uniform profiles $[0, 0.5]$ and the full circle, respectively.  For the ellipticity we used a uniform prior with width $0.02$ and centered on $0.4$ and $0.6$ for Leo K and Leo M, respectively. The use of such narrow ellipticity priors is due to an intrinsic limitation in fitting profiles to a field-of-view that is of roughly the same size as the half-light radius: the best fit tends to try to fit the field of view's ellipse, rather than the galaxy's ellipse.  By choosing an informative prior, we select out a mode that yields reasonable answers for the other parameters, at the expense of choosing an informative prior. This means our ellipticities quoted should not be viewed as well-constrained by the data, but rather as by-eye fits to provide reasonable constraints for the \emph{other} parameters.

For our distribution on the sersic shape parameter $n$, we adopted a Beta distribution prior, with $alpha=6$ and the mean set to 1. This prior has the most influence on the outcome, and was chosen to center on $n=1$ (an exponential distribution) while allowing a reasonable probability density to somewhat lower $n$, as this is common for dwarf galaxies \citep[e.g][]{McConnachie2012, Munoz2018}, while relatively quickly cutting off higher $n$. This is because the long tails of the S{\'e}rsic profile tends to produce spurious posterior peaks for data sets that do not extend to many multiples of the half-light radius. With these priors set, we used {\tt dynesty}'s dynamic nested sampling mode, with the default stopping criteria (80\% posterior-focused and 20\% evidence-focused). The resulting distributions are shown in Figure~\ref{fig:corner}, resampled for equal weights per point, and best-fitting structural parameters values are provided in Table~\ref{tab:properties}.

\begin{figure*}
\begin{center}
\includegraphics[width=0.35\textwidth]{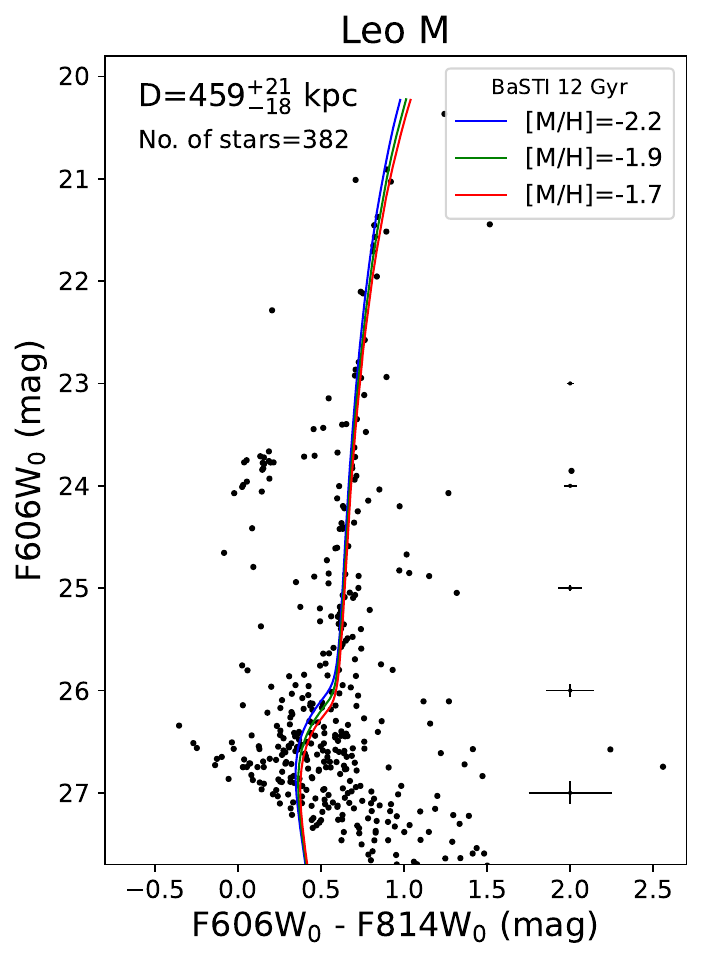}
\includegraphics[width=0.35\textwidth]{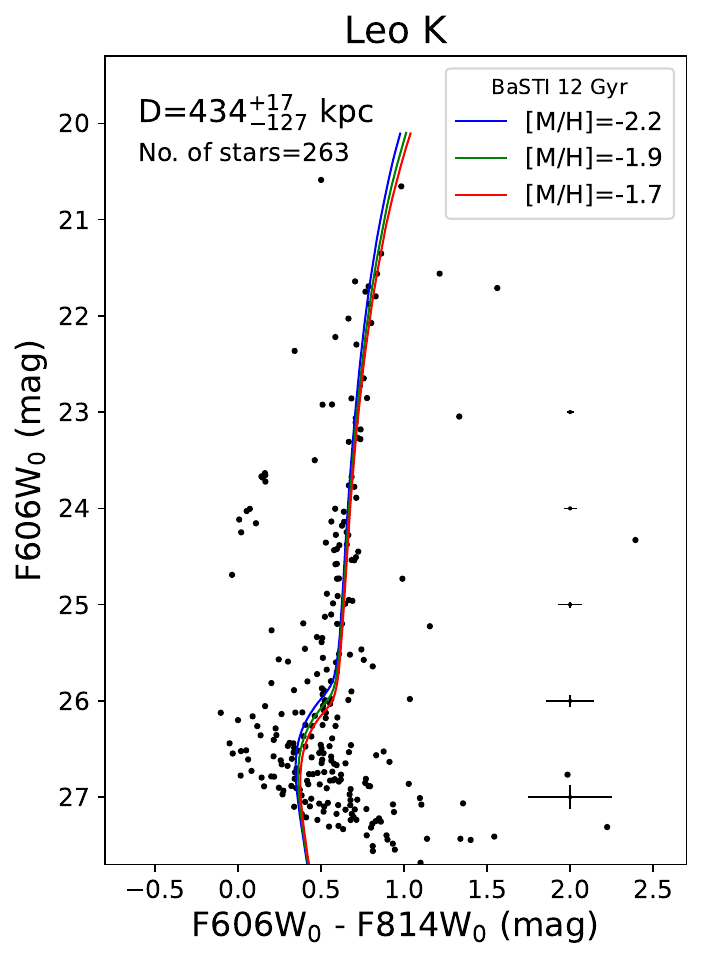}
\end{center}
\caption{Extinction corrected CMD of \leom\ (left) and \leok\ (right) with 12 Gyr BaSTI isochrones and varying [M/H] values overlaid. Representative uncertainties per magnitude are shown to the right in each panel.}
\label{fig:cmds}
\vspace{20pt}
\end{figure*}

As a visual check on the fits, Figure~\ref{fig:densprof} compares the observed and fit radial densities. The observed data are binned and shown as black points with Poissonian uncertainties. The maximum-likelihood solution is shown in red with 100 random individual draws from the posterior distributions in grey. The data and analytical profile are good matches overall.

\section{Color-Magnitude Diagrams, Distances, and Locations within the Local Group}\label{sec:cmd}
\subsection{CMDs}
Figure~\ref{fig:cmds} presents the CMDs for \leom\ and \leok\ based on the well-recovered, extinction-corrected point sources that lie within an ellipse defined by our best-fitting structural parameters out to 2$\times$ $r_h$ (see ellipses in Figure~\ref{fig:image}). Representative uncertainties per magnitude bin are also shown. Overplotted on the CMDs are 12 Gyr BaSTI isochrones spanning a metallicity range of $[M/H] = -1.7$ to $-2.2$ \citep{Hidalgo2018}; the best-fitting isochrone by eye has [M/H] $= -1.9$, in agreement within the uncertainties with the metallicity estimated from the CMD-fitting technique (see Section~\ref{sec:sfh}). 

Each CMD has a clearly identifiable, albeit sparsely populated, red giant branch (RGB), and a horizontal branch (HB) of stars. The HBs show an extension to fainter, bluer colors. Blue HB stars are generally more metal-poor than redder HB, as well as older, although the color of HB stars can also be affected by secondary parameter(s) \citep[e.g.,][and references therein]{Gallart2005}. 

Photometry of the \leom\ WFC3 parallel imaging returned only 38 sources from the full field of view that passed our quality cuts. Nearly all sources lie red-ward and fainter than the RGB of \leom\ with F606W-F814W colors greater than 1 and F606W magnitudes below 23 mag. The equivalent WFC3 data for \leok\ included only 1 source passing our quality cuts. The sparseness of the WFC3 photometric catalogs, combined with the redder colors and fainter magnitudes of the sources for \leom, suggests that the stellar catalogs presented in Figure~\ref{fig:cmds} are dominated by bona fide sources in \leom\ and \leok\ and have minimal contamination from foreground stars or background galaxies not rejected by our quality cuts.

Tables~\ref{tab:LeoM_catalog} \& \ref{tab:LeoK_catalog} provide the coordinates and HST photometry of the final stellar catalogs for \leom\ and \leok, respectively. We include the first few rows for context; the full catalogs can be downloaded in machine readable format from the online version of the paper.

\subsection{The Distance to the Galaxies}\label{sec:distance}
We determined the distances to the galaxies based on the luminosity of the HB feature in the CMDs following a multi-step procedure. As the HB feature is calibrated as a standard candle distance indicator in the $V$-band, where the brightness of HB stars is nearly constant, we first converted the photometry from the ACS filter system to the Johnson system using the transformations from \citet{Sirianni2005}. We then measured the overall HB luminosity using a maximum likelihood approach that fits a parametric function to the $V$-band magnitudes of the stars in the HB region, taking into account photometric uncertainties and completeness determined from the artificial star tests. 

\begin{figure}
\begin{center}
\includegraphics[width=0.23\textwidth]{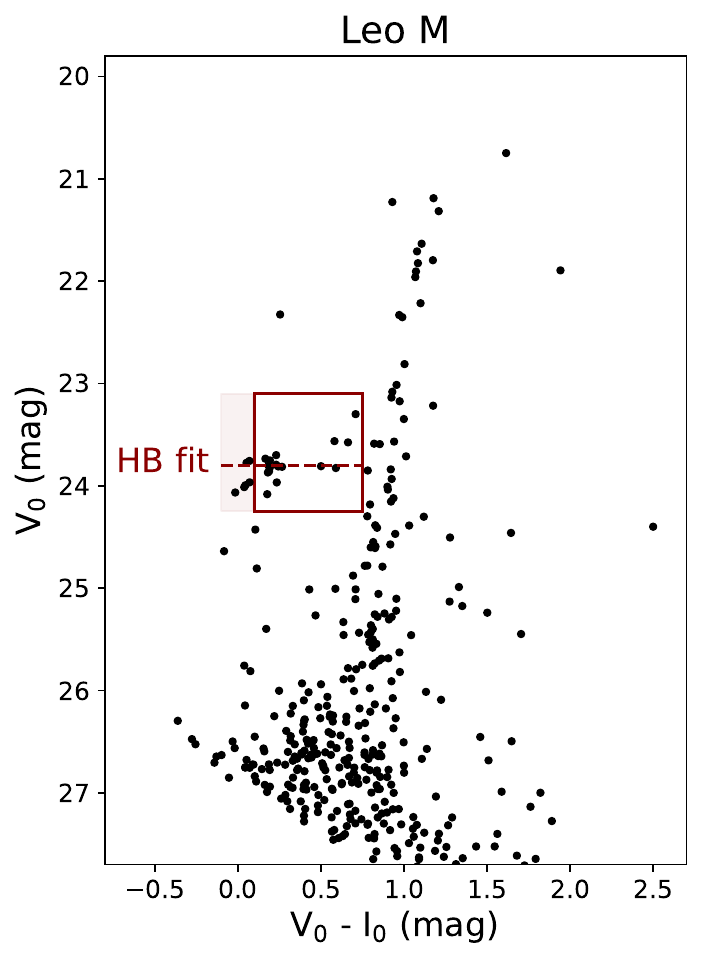}
\includegraphics[width=0.23\textwidth]{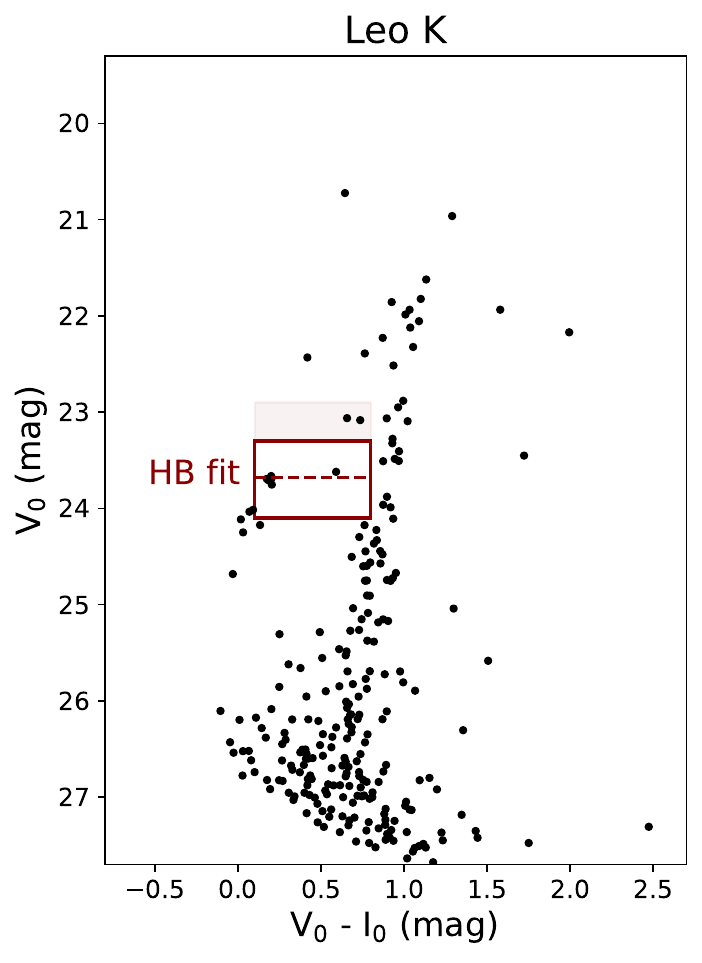}
\end{center}
\caption{CMDs of \leom\ (left) and \leok\ (right) with the HST photometry transformed to the Johnson $V,I$ system. The luminosity of the HB feature used for the distance measurement is shown as a dashed line and was determined from the stars that lie within the red boxes. Because the HBs are sparsely populated, we iterated over different color/magnitude ranges, highlighted by the shaded regions, to conservatively estimate uncertainties on the HB fit. See text in Section~\ref{sec:distance} for details.}
\label{fig:cmds_VI}
\end{figure}

Figure~\ref{fig:cmds_VI} show the CMDs of the stars transformed to the Johnson V, I system where the HB is is not as strongly sloped as in the ACS F606W filter, although a turn-down at bluer magnitudes is still seen. For the distance determination, we avoid introducing a bias from these blue, fainter HB stars by restricting our fits to stars in the HB region redward of $V-I$ of  0.1 mag. We also exclude sources with $V-I$ colors greater than 0.75 mag for \leom\ and 0.8 mag for \leok\ as these are likely red clump or RGB stars. We chose these color limits guided by the distribution of sources in the CMDs and the colors of HB stars in the latest stellar evolution models \citep[e.g., BaSTI, PARSEC, MIST;][]{Hidalgo2018, Bressan2012, Choi2016}; the regions used for these fits are highlighted by solid boxes in Figure~\ref{fig:cmds_VI}. However, because the HBs are sparsely populated, the fits can be impacted by the inclusion or exclusion of just a few sources. Thus, we iterated over the regions included in the fits to conservatively estimate our uncertainties on the distances. Specifically, for \leom, we extended the color range to $V-I=-0.1$ mag, highlighted in Figure~\ref{fig:cmds_VI} by the shaded region. For \leok, we fit for the HB using only the brighter stars close to the RGB that lie within the box but above a $V-$band mag of 23.3, shown by the shaded region in the CMD. For both the lower and upper uncertainty on the fits, we adopt the uncertainty from the maximum likelihood fit of the HB feature or the range in distances found by iterating over the different color-magnitude ranges, whichever was larger. 

We find the best-fitting HB luminosity to $M_V= 23.79^{+0.41}_{-0.04}$ mag for \leom\ and 23.65$^{+0.12}_{-0.55}$ mag for \leok. To convert to a distance, we adopt the HB metallicity-dependent distance calibration of \citet{Carretta2000}. We assume an [Fe/H] value for the HB stars of $-1.9\pm0.1$ for both galaxies based on the isochrones overlaid in Figure~\ref{fig:cmds_VI}. Note that the metallicity dependency of the calibration is quite modest. If we assumed a lower value of [Fe/H] $= -2.5$, which is closer to the value expected based on the dwarf galaxy luminosity-metallicity relation \citep{Kirby2013}, the resulting distances would still be within the uncertainties of our calculations. Our final distance modulii are 23.31$^{+0.10}_{-0.09}$ mag and 23.19$^{+0.08}_{-0.64}$ mag, for \leom\ and \leok, respectively, corresponding to distances of 459$^{+21}_{-18}$ kpc and 434$^{+17}_{-127}$ kpc. The uncertainties on the distances include an assumed uncertainty of 0.1 dex on the metallicity, the uncertainties on the calibration provided in \citet{Carretta2000}, and, as stated above, the uncertainties from the maximum likelihood fit of the HB feature or the range in distances found by iterated over our color ranges, whichever was larger. 

\begin{table*}
\begin{center}
\caption{ HST Photometry of Resolved Stars in \leom}
\label{tab:LeoM_catalog}
\end{center}
\begin{center}
\vspace{-15pt}
\hspace{-1in}
\begin{tabular}{ccccccccccc}
\hline 
\hline 
RAh	& RAm	& RAs	& DEC	& DECd	& DECm	& DECs	& F606W	& F606W err	& F814W	& F814W err \\
(hr)	& (min)	& (s)	&  sign	& ($^{\circ}$)	& (\arcmin)	& (\arcsec)	& (mag)	& (mag)	& (mag)	& (mag) \\
\hline
11 & 5 & 23.48016 & + & 25 & 19 & 16.87440 & 20.406 & 0.002 & 19.148 & 0.002 \\
11 & 5 & 17.10192 & + & 25 & 20 & 59.25120 & 20.949 & 0.003 & 20.038 & 0.003 \\ 
11 & 5 & 22.15800 & + & 25 & 20 & 30.81480 & 21.068 & 0.003 & 20.133 & 0.003 \\ 
11 & 5 & 23.46120 & + & 25 & 19 & 17.10120 & 21.485 & 0.004 & 19.954 & 0.003 \\ 
11 & 5 & 23.35032 & + & 25 & 21 & 17.35200 & 21.051 & 0.003 & 20.328 & 0.004 \\ 
11 & 5 & 20.98824 & + & 25 & 20 & 42.07200 & 21.412 & 0.004 & 20.556 & 0.004 \\ 
11 & 5 & 19.33848 & + & 25 & 20 & 47.94000 & 21.494 & 0.004 & 20.659 & 0.004 \\ 
11 & 5 & 20.31720 & + & 25 & 20 & 59.70120 & 21.556 & 0.004 & 20.648 & 0.004 \\
\hline
\hline              
\end{tabular}
\end{center}
\tablecomments{Sample of the stellar catalog for \leom\ based on the point sources passing our quality cuts and that lie within $2\times r_h$. The full catalog in machine readable format can be access via the online version of the manuscript.}
\end{table*}

\begin{table*}
\begin{center}
\caption{ HST Photometry of Resolved Stars in \leok}
\label{tab:LeoK_catalog}
\end{center}
\begin{center}
\vspace{-15pt}
\hspace{-1in}
\begin{tabular}{ccccccccccc}
\hline 
\hline 
RAh	& RAm	& RAs	& DEC	& DECd	& DECm	& DECs	& F606W	& F606W err	& F814W	& F814W err \\
(hr)	& (min)	& (s)	&  sign	& ($^{\circ}$)	& (\arcmin)	& (\arcsec)	& (mag)	& (mag)	& (mag)	& (mag) \\
\hline
9 & 24 & 7.03056 & + & 16 & 300 & 58.96800 & 20.721 & 0.003 & 19.713 & 0.003 \\
9 & 24 & 2.99376 & + & 16 & 300 & 49.22640 & 20.654 & 0.003 & 20.129 & 0.003 \\
9 & 24 & 9.35736 & + & 16 & 300 & 14.20560 & 21.627 & 0.005 & 20.388 & 0.004 \\
9 & 24 & 4.20408 & + & 16 & 300 & 35.29800 & 21.419 & 0.004 & 20.532 & 0.004 \\
9 & 24 & 6.56520 & + & 16 & 300 & 28.06200 & 21.629 & 0.005 & 20.767 & 0.005 \\
9 & 24 & 7.51776 & + & 16 & 300 & 19.63440 & 21.708 & 0.005 & 20.978 & 0.005 \\
9 & 24 & 4.33968 & + & 16 & 300 & 42.74640 & 21.759 & 0.005 & 20.948 & 0.005 \\
9 & 24 & 6.03456 & + & 16 & 300 & 47.11320 & 21.815 & 0.006 & 21.023 & 0.005 \\
\hline
\hline              
\end{tabular}
\end{center}
\tablecomments{Sample of the stellar catalog for \leok\ based on the point sources passing our quality cuts and that lie within $2\times r_h$. The full catalog in machine readable format can be access via the online version of the manuscript.}
\end{table*}

\subsection{Location within the Local Group}
\begin{figure*}
\begin{center}
\includegraphics[width=0.98\textwidth]{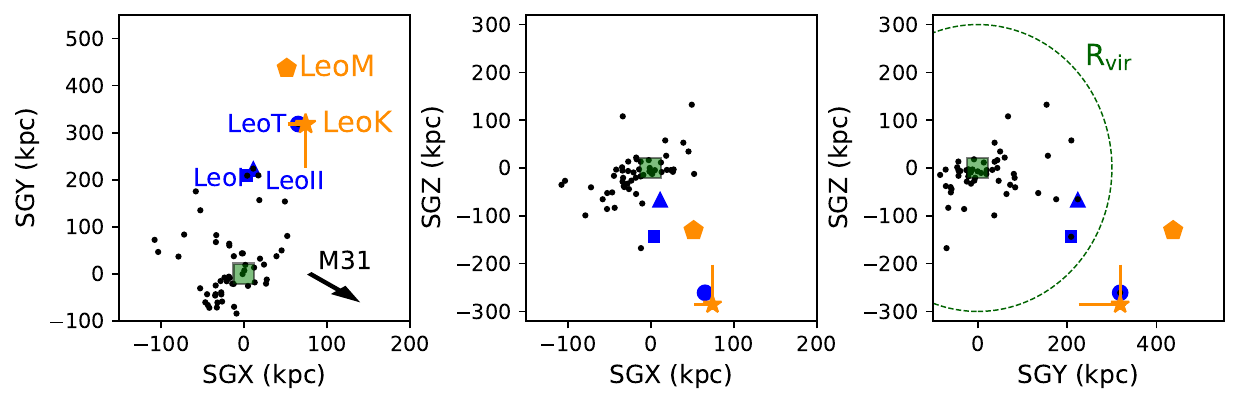}
\end{center}
\caption{Known systems in SG coordinates within 500 kpc of \leom\ and \leok\ based on the updated compilation of galaxies from \citet{McConnachie2012}. The MW is located at the SG origin, marked by a green square. \leom\ (orange pentagon) and \leok\ (orange star) are located outside an assumed 300 kpc virial radius of the MW (green dotted circle in the right panel) and below the Galactic plane (SGZ$=$0); uncertainties in the positions of the galaxies are based on distance uncertainties. The majority of black points are satellites of the MW. We highlight the three nearest known neighbors to both galaxies, Leo~I, Leo~II, and Leo~T, with a blue square, triangle, and circle, respectively.}
\label{fig:SG}
\end{figure*}

Figure~\ref{fig:SG} uses the coordinates and the secure distances to \leom\ and \leok\ to locate the galaxies within the Local Group relative to other known systems. We use the Supergalactic coordinate system, with the MW located approximately at the origin (green square), to visualize the 3-dimensional distribution of galaxies in the Galactic neighborhood. Only galaxies within 500 kpc of \leom\ and \leok\ are included in the plot. \leom\ is shown as an orange pentagon and is located $459^{+21}_{-18}$ kpc from the MW; \leok\ is shown as an orange star and located $434^{+17}_{-127}$ kpc from the MW. Both galaxies are outside an assumed MW virial radius of 300 kpc (shown as a dotted green circle in the last panel). 

We highlight the position of the 3 nearest known neighbors in blue symbols, namely Leo~I, Leo~II, and Leo~T, with 3-D separation ranging from $\sim100-250$ kpc. The exception is Leo~T which lies $\sim$ 26 kpc from \leok. This close separation is currently larger than the virial radius of either galaxy, but suggests that the two may have interacted in the relatively recent past. We return to this in Section~\ref{sec:conclusions}. For clarity, all other galaxies, the majority of which are satellites of the MW, are shown as smaller black circles.

\section{The Star Formation Histories}\label{sec:sfh}
\subsection{SFH Methodology}
The SFH was measured using the CMD-fitting technique {\sc match} \citep{Dolphin2002}. Briefly, {\sc match} generates synthetic photometry of simple stellar populations with different ages and metallicity using stellar evolutions libraries and an assumed an initial mass function (IMF), and adopting a set of galaxy-specific parameters. These synthetic CMDs are convolved with the photometric uncertainties and completeness function determined from the artificial star tests, combined with different weights, and iteratively compared to the observed CMD until the best fit is found using a Poisson likelihood function. 

For the SFHs of \leom\ and \leok, we assumed a Kroupa IMF \citep{Kroupa2001} and a binary fraction of 0.35 with flat secondary mass ratio distribution.\footnote{SFH solutions have been shown to be quite stable over a range of assumed binary fractions \citep{Monelli2010}. We adopt a fraction of 0.35 as that is frequently used in SFH work and provides a more direct comparison with works in the literature.} We adopted a foreground extinction value of $A_V=0.045$ and 0.073 mag, respectively \citep{Schlafly2011}. We fixed the distances in the fits based on our distance measurements from the HB stars and require that the metallicity monotonically increase with time. While contamination from foreground stars is expected to be minimal based on small number of sources photometrically recovered from the WFC3 parallel fields, we nonetheless included a model of possible foreground contamination in the fits using Trilegal simulations \citep{Girardi2012}. Statistical uncertainties were estimated using a hybrid Markov Chain Monte Carlo approach \citep{Dolphin2013}. Systematic uncertainties were estimated via 50 Monte Carlo simulations that apply shifts in luminosity and color and re-fit for the SFH \citep{Dolphin2012}.  The SFHs were reconstructed using three stellar evolution libraries, namely BaSTI, PARSEC, and MIST. In addition to the conservative systematic uncertainties estimated from the Monte Carlo simulations, the range in solutions from the 3 libraries brackets the range of likely SFH solutions providing an additional indication on the systematic uncertainties of the fits. Internal extinction was estimated to be zero by iteratively fitting for the SFH using each stellar library and varying levels of extinction until the lowest fit value was found for each galaxy. 

\subsection{Best-Fitting SFHs}
\begin{figure*}
\begin{center}
\includegraphics[width=0.48\textwidth]{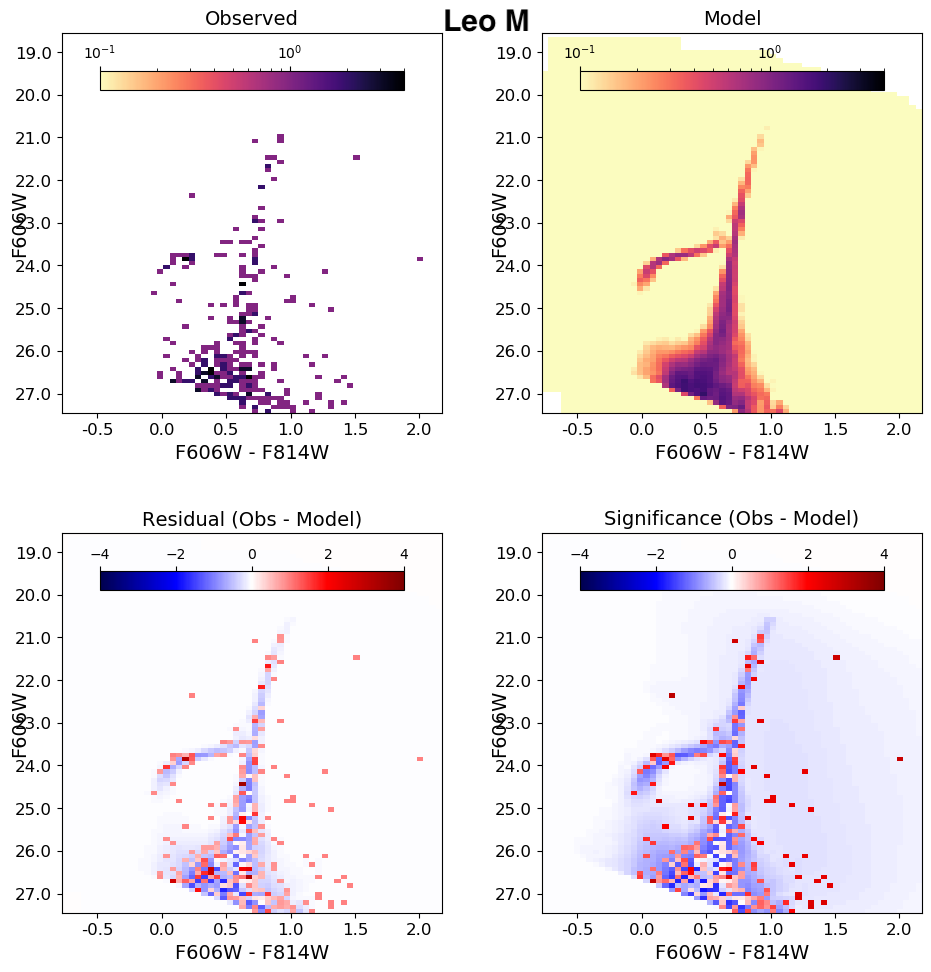}
\includegraphics[width=0.48\textwidth]{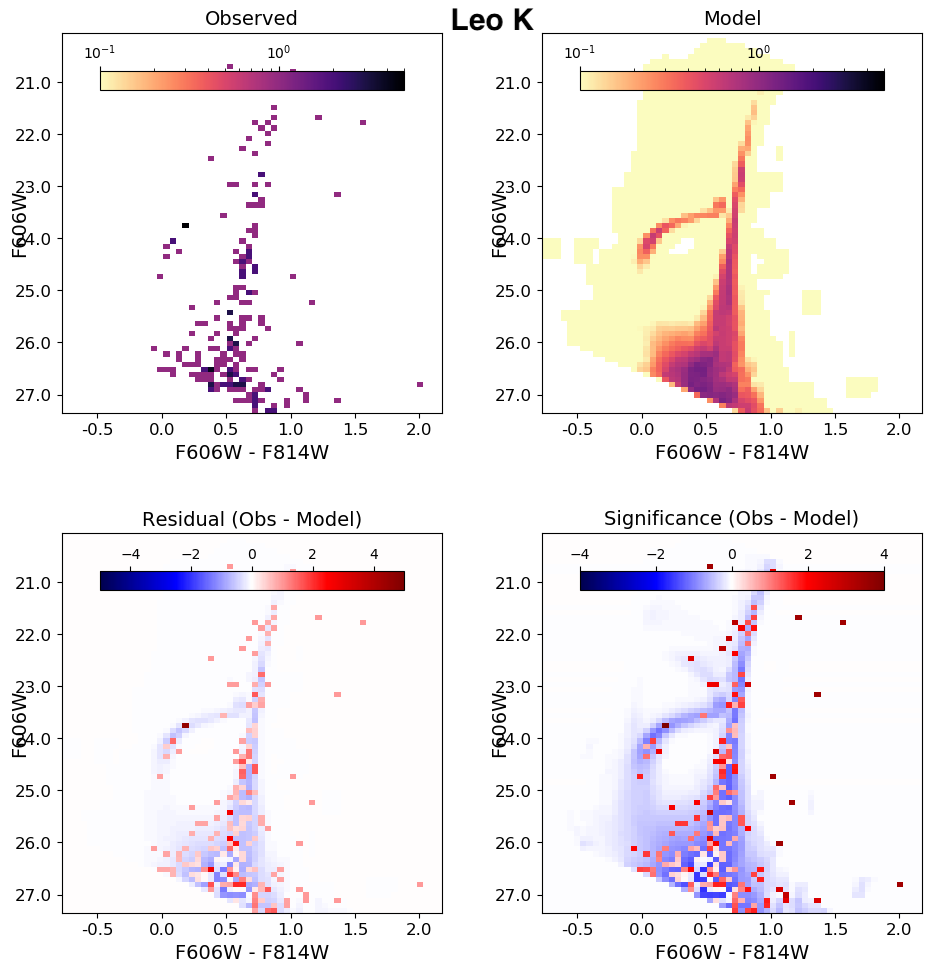}
\end{center}
\caption{Example of the quality of fit to the observed CMD using the BaSTI stellar library shown in 4-panel plots for \leom\ (left) and \leok\ (right). For each plot series: top left is the observed CMD; top right is the modelled Hess diagram used in the fit; bottom left is a simple residual Hess diagram (data $-$ model); bottom right is the residual significance Hess diagram where the pixels are weighted by the variance. The checkerboard pattern in the residual significance plots indicates there are no major residuals and the modelled Hess diagram is a good fit to the data.}
\label{fig:residuals}
\end{figure*}

Figure~\ref{fig:residuals} shows the quality of the SFH fits based on the BaSTI stellar library. For each galaxy, we present in Hess diagram format the observed CMD (top left), the modelled CMD based on the BaSTI library (top right), the residuals between the observed and modelled CMDs (bottom left), and the significance of the residuals or the observed $-$ model weighted by the variance in each Hess bin (bottom right). Overall, the modelled CMDs are well-matched to the data, with no clear trends seen in the weighted residuals. We note the largest difference between the modelled CMDs is seen in the fainter blue HB stars, which are present in the BaSTI synthetic CMD but largely absent in the PARSEC and MIST synthetic CMDs (not shown). 

Figure~\ref{fig:sfh} presents the best-fitting SFHs for \leom\ and \leok. For each galaxy, the left panel shows the SFH fit using the BaSTI stellar library (red lines) with uncertainties (shaded orange) that include both statistical and systematic uncertainties; the right panel compares the BaSTI results to the SFHs derived using the PARSEC (blue) and MIST (black) models with statistical uncertainties only. For both galaxies, the three models are in very good agreement with each other. We adopt the BaSTI SFH solutions for the downstream analysis in the paper. This also allows for a direct relative comparison with the final adopted SFH for \peg, which was also derived using the BaSTI models. The gray shaded vertical region shows the approximate timescale of reionization.

\subsection{Star-Formation Timescales}
For each galaxy, we determined the star formation ``quenching timescale'', or the time after which the system has formed relatively little of its stellar mass. In our previous work on \peg, we adopted the \tninety\ metric as the quenching timescale, defined as the lookback time by which 90\% of the stellar mass has been formed. However, as pointed out in \citet{Savino2023}, the \tninety\ metric is a better measure of quenching for the well-populated CMDs of more massive galaxies. Given the low star counts in the CMDs of ultra-faint dwarfs, \tninety\ may be subject to spurious sources impacting the estimate of late time star formation. Thus, we adopt the more conservative lookback time at which 80\% of the stellar mass is formed, or \teighty, as our quenching timescale. 

The value of \teighty\ and the uncertainty are found by interpolating the SFH solutions from the BaSTi library. For \leom\ and \leok\ we find similar quenching timescales of \teighty\ $= 10.6^{+2.2}_{-1.1}$ Gyr and $12.8^{+0.1}_{-4.2}$ Gyr, respectively. For direct comparison with \peg, we also calculate \teighty\ based on the SFH from the BaSTI library in \citet{McQuinn2023} to be $8.0^{+4.7}_{-1.6}$ Gyr, slightly older than the reported \tninety\ value of $7.4^{+2.2}_{-2.6}$ Gyr. 

\subsection{Luminosities and Stellar Masses of the Galaxies}
We estimated the total luminosity and stellar mass of the galaxies by generating synthetic stellar populations based on the best-fitting SFH and distance measurement using a Monte Carlo approach. Specifically, we generated synthetic photometry adjusted to a distance drawn from the distance measurement and uncertainties. We then selected stars from this population until we reached the number of stars, $N_*$, that matched the number of {\em observed} stars in our previously chosen CMD limits used in fitting the structural parameters. This synthetic stellar catalog was corrected for extinction and distance, and the F606W magnitudes are converted to $V$-band magnitudes adopting an [Fe/H] value of $-1.9$ and using the bolometric corrections from \citet{Chen2019}. The final integrated magnitudes determined for \leom\ and \leok\ are $M_V$ are $-5.76^{+0.15}_{-0.16}$ mag and $-4.83^{+0.81}_{-0.28}$ mag, respectively. We also list these values in Table~\ref{tab:properties}.

\begin{figure}
\includegraphics[width=0.47\textwidth]{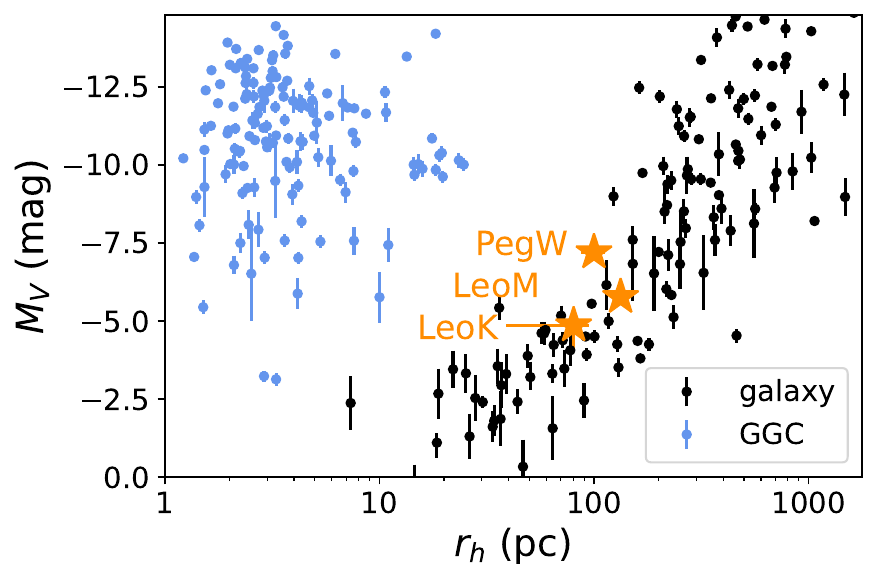}
\caption{$M_V$ magnitudes vs.\ half-light radius ($r_h$) in units of pc for low-mass galaxies and Galactic globular clusters (GGCs). Overlaid as large orange stars are the locations of \peg, \leom, and \leok. The galaxies are consistent with the properties of known local dwarfs with comparable luminosity, although slightly more compact, and are considered an ultra-faint dwarf galaxy based on the definition from \citet{Simon2019}. }
\label{fig:Mv_rh}
\vspace{0.2in}
\end{figure}

Figure~\ref{fig:Mv_rh} uses the estimated $M_V$ and $r_h$ values to compare the two galaxies with \peg, other low-mass galaxies (black points) from \citet{McConnachie2012}, and the compilation of Galactic globular clusters (GGCs; blue points) from \citet{Baumgardt2020}. Both galaxies lie within the distribution expected for low-mass galaxies, and are clearly separate from the GGC population. Similar to \peg, \leom\ and \leok\ lie on the more compact side of the galaxy distribution, which likely reflects that more compact galaxies are more readily detected via stellar over-density search techniques. 

The total mass of the stars was estimated using two approaches. First, we summed the masses of stars drawn in each Monte Carlo iteration used for the luminosity calculation, which gave stellar masses estimates of \leom\ and \leok\ of \mstar\ $=2.1^{+0.3}_{-0.2}\times 10^4$ \msun\ and $9.7^{+1.8}_{-5.5}\times 10^3$ \msun, respectively. Second, we estimated the present-day stellar mass directly from the SFH results. Assuming mass limits on a Kroupa IMF of $0.1-100$ \msun, and adopting a recycling fraction of 41\% for gas returned to the ISM from the stars \citep{Vincenzo2016}, we find \mstar\ $=1.8^{+0.3}_{-0.2}\times 10^4$ \msun\ and $1.2\pm0.2\times 10^4$ \msun. The stellar masses from the two methods are in good agreement, but note that the masses determined from the SFH fits also include the mass of stellar remnants. For our final values, we adopt the stellar masses based on the SFH fits.

\begin{figure*}
\begin{center}
\includegraphics[width=0.48\textwidth]{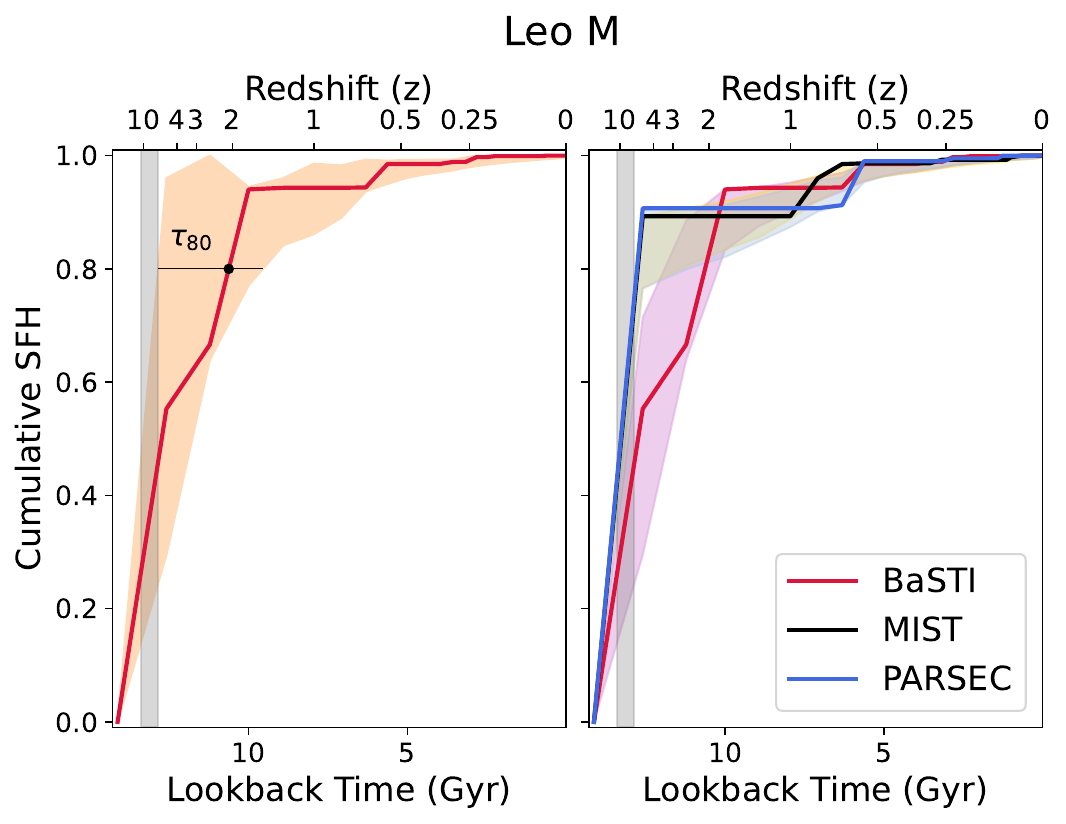}
\includegraphics[width=0.48\textwidth]{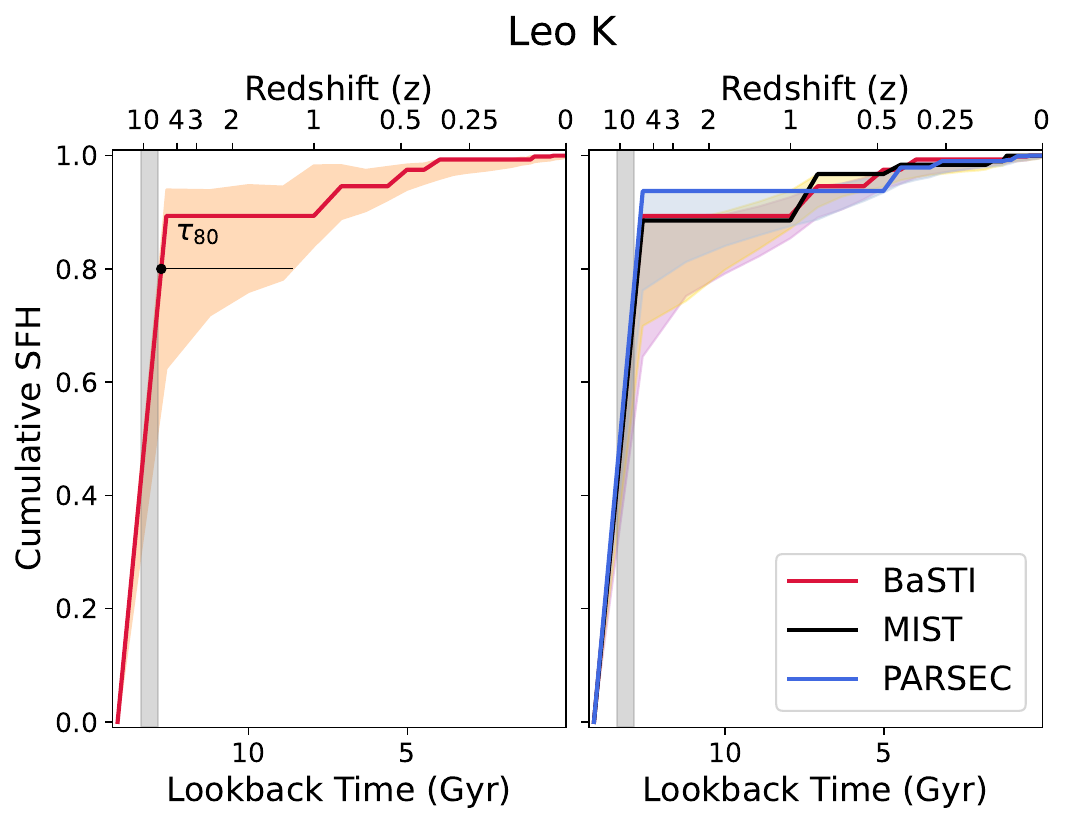}
\end{center}
\caption{For each galaxy we show the best-fitting SFH solution with the BaSTI models (left panels). Uncertainties include both statistical and systematic uncertainties. The quenching timescale, \teighty, is marked. Also shown as the best-fitting SFH solutions using the BaSTI, MIST, and PARSEC models with statistical uncertainties only shaded in pink, yellow, and blue, respectively (right panels). The gray shaded vertical region shows the approximate timescale of reionization.}
\label{fig:sfh}
\end{figure*}

\section{Exploring the Intertwined Impacts of Reionization and Environment}\label{sec:discuss}
While early quenching of ultra-faint dwarfs is typically attributed to reionization, as described in Section~\ref{sec:intro}, distinguishing the effects of environmental and reionization quenching at this low mass scale ($\lesssim 10^4$ \msun) can be challenging. Not only are the orbital histories of galaxies difficult to constrain observationally, simulations have shown that the infall time for satellite ultra-faint dwarfs can span a very wide range, including a sizable fraction of satellites falling into the host systems before 8 Gyr ago \citep{Simpson2017,Pan2022}. Because \leom\ and \leok\ are both sitting outside of the virial radius of the Milky Way, despite the current lack of any measured galaxy velocities, there is a higher probability that they have later infall times \citep{Simpson2017}. The presumably more isolated environment at early times simplifies interpretations: the rapid quenching of \leom\ and \leok\ is consistent with the combined effect of stellar feedback and reionization where gas internal to a galaxy is both heated and ejected while gas accretion onto the galaxy is halted. 

Interestingly, the ultra-faint dwarf \peg\ has an extended SFH lasting several Gyr, which is inconsistent with reionization quenching. \peg\ is more massive than \leom\ and \leok, but still below the mass scale ($10^5$ \msun) where galaxies are typically predicted to be quenched by reionization in simulations. As suggested in \citet{McQuinn2023}, \peg\ was most likely quenched via environmental processes (i.e., via a previous fly-by with M31). This late quenching timescale of \peg\ is in contrast not only to our findings for \leom\ and \leok, but for nearly all ultra-faint dwarfs satellites of the MW and M31 with detailed SFHs, including satellites that have present-day masses greater than \peg\ \citep[e.g.,][]{Brown2014, Wetzel2014, Savino2023}.

If \leom\ and \leok\ were indeed quenched by reionization while \peg\ was quenched later by environmental effect, this apparent difference may naively be attributed to a reionization quenching mass scale between 1 and 6 $\times 10^4$\msun, slightly lower than recent simulation predictions \citep[e.g.,][]{RodriguezWimberly2019, Rey2020}. However, the timing and degree of impact of reionization on a very low-mass halo may vary as a function of the large-scale environment as reionization proceeded in a non-homogenous manner, with galaxies creating ionizing bubbles that grew in radius with time. Recent JWST observations of the most distant Ly$\alpha$-emitting galaxies strongly suggest that reionization was spatially inhomogenous \citep{Leonova2022}. Size estimates of the ionizing bubbles surrounding galaxies during the epoch of reionization, while still uncertain, are consistent with a radius of $\sim0.5$ physical Mpc at $z \gtrsim 9$ \citep{Hayes2023,Umeda2023} (the bubble radius may be smaller at even higher $z$). \citet{Witstok2023} find bubble sizes of 0.1$\sim$1 pMpc at 5.8$<$$z$$<$8 with the bubbles embedded in a completely neutral IGM. Given these results, it is possible that low mass galaxies at different distances from a more massive galaxy may have been enveloped by the ionized bubbles over different timescales \citep{Kim2023}. 
Hence, it is conceivable that \leom\ and \leok, as well as most of the quenched ultra-faint dwarf satellites, were within the MW-M31 ionizing bubbles at early times, while \peg\ was farther afield and felt the impact of reionization at a somewhat later time, resulting in a very different SFH. 

\section{Conclusions}\label{sec:conclusions}
We report the discovery of two ultra-faint dwarf galaxies \leom\ and \leok. Both galaxies were observed with the HST as part of GO-16916 (PI McQuinn), which also targeted the ultra faint dwarf galaxy, \peg. These three galaxies are unique in that they are ultra faint dwarfs located outside the halo of a massive galaxy and, therefore, can provide new constraints on how very low-mass halos assemble their stellar mass in a less dense, and presumably less complicated, environment than the satellites of the MW, M31, and the Clouds. Our main findings are:

\begin{itemize}
\item \leom\ and \leok\ have very low stellar masses (\mstar\ $=1.8^{+0.3}_{-0.2}\times10^4$; $1.2\pm0.2\times10^4$ \msun), and are very faint ($M_V=-5.77^{+0.15}_{-0.16}$; $-4.86^{+0.83}_{-0.29}$mag). Both galaxies are located outside the virial radius of the MW at 459$^{+21}_{-18}$ and 434$^{+17}_{-127}$ kpc, respectively, based on the luminosity of their HB stars (Figures~\ref{fig:cmds} $-$ \ref{fig:SG}). Their integrated $V-$band magnitudes and half-light radii are consistent with the properties of other local low-mass galaxies, although both are somewhat more compact than the majority of known systems (Figure~\ref{fig:Mv_rh}). Based on our analysis, \leom\ and \leok\ are ultra faint dwarfs located within the Local Group but are not currently within the virial radius of a more massive host. 

\item \leok\ is located $\sim26$ kpc from the low-mass galaxy Leo~T. Previously, \citet{Adams2018} noted subtle signs of an interaction in the \hi\ emission of Leo~T from deep Westerbork Telescope observations of the 21 cm line, including a truncation of the \hi\ emission on the western edge and an offset of the peak \hi\ emission to the south from the optical center. These authors suggest the \hi\ features may be related to a past interaction with the circumgalactic medium of MW, despite Leo~T being 420 kpc from the Galaxy. \leok\ is located to the west of Leo~T in the same direction of the \hi\ truncation, offering an alternative interpretation. It is possible the subtle features in the \hi\ morphology of Leo~T were due to the interaction of these two small systems, rather than with the MW. 

\item The SFHs of \leom\ and \leok\ (Figure~\ref{fig:sfh}) based on deep HST imaging show that both galaxies formed the majority of their stellar mass at early times. Specifically, \leom\ and \leok\ were quenched $10.6^{+2.2}_{-1.1}$ and $12.8^{+0.1}_{-4.2}$ Gyr ago, respectively, based on when each system formed 80\% of its stellar mass (\teighty). Similar to what has been noted for many other ultra-faint dwarfs, the SFHs for both galaxies suggest that a small fraction of the stars were formed at intermediate times.
\end{itemize}

We have also compared the results on \leom\ and \leok\ with those on the slightly more massive ultra-faint dwarf \peg\ that does not appear to have been quenched early by reionization \citep{McQuinn2023}. While a very small sample, our findings suggest that the mass scale for reionization to quench ultra-faint dwarfs may be lower than typically predicted and/or it may depend on the distance to a more massive system at early times. A large uncertainty in the current interpretation is the mass of the dark matter halos hosting the galaxies and their velocities, which could be constrained via follow-up spectroscopy.

\acknowledgments
We would like to acknowledge Eddie Schlafly for helpful discussions regarding the structural parameter modeling. We also thank the anonymous referee for helpful comments that improved this paper. Support for this work was provided by NASA through grant No.\ HST-GO-16916 awarded by the Space Telescope Science Institute, which is operated by the Association of Universities for Research in Astronomy, Incorporated, under NASA contract NAS5-26555. D.S.\ and M.R.B.\ are supported by DOE grant DOE-SC0010008. This research has made use of NASA Astrophysics Data System Bibliographic Services, adstex\footnote{https://github.com/yymao/adstex}, and the arXiv preprint server. 

\facilities{Hubble Space Telescope}
\software{This research made use of {\tt DOLPHOT} \citep{Dolphin2000, Dolphin2016}, {\tt MATCH} \citep{Dolphin2002, Dolphin2012, Dolphin2013}, HST drizzlepac \citep[v3.0;][]{Hack2013, Avila2015}, {\tt emcee} \citep{Foreman-Mackey2013},  {\tt dynesty} \citep{dynesty1, dynesty2}, Scipy \citep{scipy}, and Astropy,\footnote{http://www.astropy.org} a community-developed core Python package for Astronomy \citep{astropy:2013, astropy:1801.02634, astropy:2022}.}

\vspace{0.6in}
\renewcommand\bibname{{References}}
\bibliographystyle{apj}
\bibliography{ms_revised}

\end{document}